# Sensing enhancement of a Fabry-Perot THz cavity using switchable VO$_2$ mirrors

Gian Paolo Papari,[1,2,3,*] Anna Lucia Pellegrino[4] Graziella Malandrino[4] and Antonello Andreone[1,2,3]

[1]Dipartimento di Fisica, Università di Napoli "Federico II," via Cinthia, I-80126 Napoli, Italy

[2]CNR-SPIN, UOS Napoli, via Cinthia, I-80126 Napoli, Italy

[3]Istituto Nazionale di Fisica Nucleare (INFN), Naples Unit, via Cinthia, I-80126 Napoli, Italy

[4]Dipartimento di Scienze Chimiche, Università di Catania, and INSTM UdR Catania, viale A. Doria 6, I-95125 Catania, Italy

*Corresponding author: gianpaolo.papari@unina.it

**Abstract:**

We experimentally investigate the sensing properties of an open cavity operating in the THz regime and realized by employing as mirrors two thin vanadium dioxide (VO$_2$) films grown on silicon parallel plates and separated by a variable length. The phase transition of VO$_2$ is used to control the behavior of the system between two different responses: a high transmission mode to the incident radiation (VO$_2$ in the insulating state) and a high sensitivity to tiny changes in the cavity refractive index (VO$_2$ in the conducting state). In the first state, the low loss regime enables to adjust the cavity length and easily optimize the resonances due to the Fabry-Perot (FP) effect in the Si plates and in the cavity volume. The activation of the metallic-like state instead, by damping the FP oscillations in the plates, promotes the onset of a comb-like spectrum that can be exploited as a versatile tool for accurate sensing applications. Using both an analytical model and full-wave simulations, we estimate the device response to variation in the refractive index of the cavity volume, showing that the proposed structure can achieve sensitivity values among the highest reported for THz sensors.

## 1. Introduction

One of the most effective working principle of an optical sensor grounds on how specific inherent resonances respond to changes in the refractive index (dielectric function) of the element to be probed, usually, the environment where the sensor is inserted or is in contact with [1]. During the years, several typologies of resonators have been proposed for sensing purposes. For instance, sensors realized by using metamaterials and metasurfaces [2] [3], [4] [5], waveguides [6] and bulk cavities [7] have been reported in the recent literature. High sensitivity devices need a specific design to properly work in a targeted frequency band. Each resonance frequency is connected to a specific eigenmode whose activation can be engineered in realizing the sensor through an appropriate design of metallic/dielectric parts. Hence, a high sensitivity sensor is usually a tailored device with single aim and characteristic spectral response.

Conversely, a tunable parallel plate cavity realized with two mirrors in series can show a shiftable comb-like spectrum depending on the cavity length ($L$), that is the distance between the mirrors [8]. The transmission spectrum of such a sensor is due to the Fabry-Perot (FP) interferential mechanism [9] of waves reflected back and forth within the cavity. In transmission configuration the FP effect generates a series of maxima (FP modes) which are periodically distributed in the frequency band. As properly discussed below, the sensitivity of the cavity grows according to the refractive index of the conducting layer $n_m$ because the larger $n_m$ the higher the quality factor of the inherent resonances [9]. $L$ can be tuned to optimize the maxima distribution according to each specific target.

Owing to the individual response and very high sensitivity of each maximum, such a device based on a tunable cavity can represent a disruptive technological advancement in the panorama of THz sensors. Unfortunately, a major drawback of this type of sensor is caused by the low transmitted signal because of the unavoidable presence of conducting plates acting as mirrors. Indeed, a parallel plate cavity is typically realized depositing a metallic layer, thinner than the penetration depth, on a dielectric substrate. Considering that a 10 nm thick Au film deposited on Si enables a transmission $|\tilde{T}|$ less than 0.1 [10] and signal intensity across two films scales as $|\tilde{T}|^2$, a metallic mirror parallel plate cavity would display a very weak output, barely useful for practical applications.

A promising solution to overcome the competing effects between increased sensitivity and reduced resolution consists in controlling the refractive index of the mirrors, varying $n_m$ from a low (OFF) to a high (ON) value. In such a way by handling $L$ one can properly tune $|\tilde{T}|$ with ease when the cavity is in the OFF mode, and then switch the system in the ON mode for the sensing operation.

In this work we present the proof-of-concept of a Parallel Plate Fabry-Perot Cavity (PPFPC) with a tunable length realized using as mirrors two vanadium dioxide ($VO_2$) films deposited on Si substrates. $VO_2$ undergoes a phase transition from an insulating to a conducting electrical behavior when its temperature rises above a critical value $T_c$ close to room temperature, at around $77°C$ [11] [12] [13]. By exploiting the phase transition of $VO_2$ it is possible to switch the device from an "open" system consisting in a pair of two semi-transparent surfaces (with respect to the impinging electric field) to a "closed" structure marked by two metal-like mirrors and endowed with enhanced sensing features [14].

To focus the study on the variation of the cavity properties produced by the activation of the metal-like phase in the $VO_2$ films, the paper discusses the electrodynamic response of the PPFPC at two temperatures only, below (T = 50 °C) and above (T = 90 °C) the transition temperature of the vanadium dioxide. We will show that the quality factor of resonances excited in the parallel plate cavity by the FP mechanism more than doubles after the phase transition, promoting this simple device as an optimal candidate for accurate sensing in cost-effective applications.

The paper is divided in various parts: in Section 2 the experimental setup is described, in Section 3 the electrodynamic properties of a single plate are introduced, in Section 4 the transmission signals in time and frequency domain are presented, in Section 5 the analytical model of the PPFPC response is reported with a discussion on how to optimize its maxima, in Section 6 the device sensitivity is commented through analytical results and full-wave simulations, then final conclusions are drawn.

## 2. Experimental techniques

We employed a time domain THz system (TERA K15 by Menlo Systems®) based on photoconductive antenna technology [15] [16]. Signals are acquired in a time interval of about 250 ps providing a frequency resolution of about 4 GHz. Measurements are performed in a box purged with $N_2$ gas to minimize the absorption by water vapor. To ensure a transmitted THz signal as close as possible to a planar wave we used two TPX lenses in the optical path to collimate the beam immediately after the emitter and before the detector [17]. In Fig.1 a sketch of the optical setup is shown.

The PPFPC is realized using two high resistivity silicon plates having thickness d = (400 ± 10) μm over which a thin $VO_2$ layer, [$t$=(1.10±0.05) μm] is grown [17]. $VO_2$ films are placed face-to-face, and their relative distance $L$ is varied in the range 240 ÷ 7800 μm through a motorized linear stage.

Metal Organic Chemical Vapor Deposition (MOCVD) approach was applied for the fabrication of $VO_2$ films on 1 cm × 2 cm Si substrate, following a procedure similar to that previously reported [18]. The vanadyl-acetylacetonate $VO(acac)_2$, purchased from STREM Chemicals, was used as precursor without further purifications. The $VO_2$ films were deposited for 1 h in a customized, horizontal, hot-wall reactor, using argon (300 sccm) as carrier gas, and oxygen (150 sccm) as reacting gas. The pressure inside the reactor was kept at the value of 4 Torr through a scroll pump unit and monitored using MKS Baratron 122AAX. The Ar and $O_2$ flows were controlled using MKS 1160 flow controller units. Deposition temperature was fixed at 450°C and the precursor source was kept at 170°C for an efficient vaporization process. Each section was heated using K-type thermocouples and computer-controlled hardware with ±2°C accuracy. A comprehensive analysis of $VO_2$ film was performed through X-ray diffraction (XRD) technique and field emission scanning electron microscopy (FE-SEM) for structural and morphological features, respectively. XRD characterization was performed using a Smartlab Rigaku diffractometer in grazing incident mode (0.5°) operating at 45 kV and 200 mA, equipped with a rotating anode of Cu Kα radiation. FE-SEM images were obtained using a field emission scanning electron microscope ZEISS SUPRA 55 VP.

In Fig.2(a), the XRD diffraction pattern displays the peaks at 27.94, 37.11, 39.86, 42.28, 55.54 and 57.62° associated with the formation of a pure and crystalline monoclinic $P2_1/c$ phase (PDF file numbers 43-1051, 44-0252). In Fig. 2(b), the FE-SEM image of the $VO_2$ film shows the formation of a very homogeneous layer with plate-like grains of the order of 700 nm. The cross-sectional image, in Fig. 2(c) shows a very compact film with a thickness of about 1.10 ±0.05 μm.

To control the phase transition of the vanadium dioxide layers, the substrates are glued with an appropriate epoxy on heaters made of copper and thermally controlled. Our $VO_2$ samples undergo a phase transition at about $T_c = 78\ °C$ [11].

## 3. Properties of a single $VO_2$ film.

When the sample temperature is kept below $T_c$, the $VO_2$ layers are insulating and basically transparent to the impinging THz radiation. Instead, for $T > T_c$ each $VO_2$/Si plate behaves as a metal-like mirror.

In Fig. 3(a) the time dependent signal transmitted through a single $VO_2$ film on silicon is reported at two different temperatures, below and above $T_c$. Black curve is recorded for $T \sim 50°C$, therefore it shows features ascribable to the bare Si substrate only. On the contrary, the red curve is acquired at temperatures above $T_c$, approximately $90\ °C$. Here, clear signatures of a conducting behavior are observed. In fact, in addition to a strong attenuation in the main signal intensity, the Fabry-Perot odd reflections are reversed, because of the THz beam reflection at the $VO_2$/Si interface introducing a factor -1 in the phase response (see discussion below).

Once the refractive index of the bare Si plate $\tilde{n}_s = n_s + ik_s$ is extracted, the complex refractive index of the thin film $\tilde{n}_v = n_v + ik_v$ in the metallic phase can be easily evaluated by an appropriate comparison between the experimental $\tilde{T}_e(\omega)$ and theoretical transmissions $\tilde{T}$ [19]. Specifically, by applying the Fast Fourier

Transform (FFT) to the signal transmitted through the VO$_2$/Si sample, $\mathfrak{F}\{E_s(t)\} = \tilde{E}_s(\omega)$, and to the reference one in absence of the sample, $\mathfrak{F}\{E_r(t)\} = \tilde{E}_r(\omega)$, one can measure $\tilde{T}_e(\omega) = \tilde{E}_s(\omega)/\tilde{E}_r(\omega)$. Instead, $\tilde{T}$ is achieved by using the appropriate combination of Fresnel coefficients and propagation factors yielding [20]

$$\tilde{T}(\omega) = 2\frac{\tilde{n}_f(\tilde{n}_a+\tilde{n}_s)}{(\tilde{n}_a+\tilde{n}_v)(\tilde{n}_v+\tilde{n}_s)}\exp\left\{-i\omega\frac{t}{c}(\tilde{n}_v-\tilde{n}_a)\right\}\widetilde{FP}(\omega) \qquad (1)$$

In the above formula, $\tilde{n}_a = 1$ is the complex refractive index of the air, $c$ is the speed of light in vacuum, and $\widetilde{FP}(\omega)$ is the Fabry-Perot term [21] related to the multiple reflections inside the film:

$$\widetilde{FP}(\omega) = \frac{1}{1-\left(\frac{\tilde{n}_v-\tilde{n}_a}{\tilde{n}_v+\tilde{n}_a}\right)\left(\frac{\tilde{n}_v-\tilde{n}_s}{\tilde{n}_v+\tilde{n}_s}\right)\exp\left\{-i\omega\frac{2t}{c}\tilde{n}_f\right\}} \qquad (2)$$

This term introduces an oscillatory behavior in the transmission response whose peaks (maxima) define the FP modes within the slab. Maxima are distributed in frequency with a periodicity $\Delta f_{FP_v} = c/(2n_v t)$, basically related to the back-and-forth trip of the "effective optical length" $n_v t$. $\tilde{n}_v$ is calculated through the application of a computational technique aimed to minimize the difference $\tilde{T}_e - \tilde{T}$ varying $n_v, k_v$ [19]. From here the VO$_2$ conductivity is achieved using the formula $\sigma = 2\,\varepsilon_0\omega\,n_v k_v$ where $\varepsilon_0$ is the permittivity in vacuum. The net change in the conductivity of the VO$_2$ film is shown in Fig. 3(b) for the two set temperatures. At $T = 50\,°C$ the conductivity is of the order of 2 S/cm with a flat frequency dependence, whereas at $T = 90\,°C$ $\sigma$ displays a strongly dispersive behavior with values ranging between 100 and 400 S/cm over almost a frequency decade, in accordance with previous reports [13]. Inaccuracy in the extraction of the sample electrodynamic parameters is mostly determined by the uncertainty on the film thickness value, giving an overall conductivity error lower than 10% [19].

### 4. The VO$_2$ cavity

The time dynamics of the electric field signal $E_c(t)$ transmitted through the PPFPC is basically ascribable to the coherent superposition among FP reflections generated within a single VO$_2$/Si plate with those taking place in between the two plates.

For the sake of clarity, we first explain the basic electrodynamics ruling the PPFPC time response when the VO$_2$ layers are in the insulating state (T = 50 °C). In this case the transmission is substantially driven by the electrodynamics of the two Si substrates. $E_c(t)$ curves recorded for different values of L are shown in Fig. 4(a), and the system can be schematized using four interfaces only, labelled I1-I4 as reported in Fig. 4(b).

Most of the phenomenology here pertains the concurrence between the periodicity in the frequency response of the cavity defined by the VO$_2$ layers and filled with a medium of refractive index $n_c$, $\Delta f_{FP_c} = c/(2Ln_c)$, and the analogous mechanism produced by the FP oscillations inside each plate, $\Delta f_{FP_s} = c/(2dn_s)$. Controlling $L$ and $n_c$ one can tune the frequency spacing of the cavity modes to be larger or smaller than the FP modes inside the substrates. In what follows it is assumed $n_c = n_a = 1$ but it will be replaced by a variable quantity in the last Section to estimate the sensitivity of the cavity.

Analyzing how the time delay between different peaks changes varying L is extremely useful to understand the dynamics of THz pulse reflections on the cavity interfaces I2 and I3. In Fig. 4(a) the black curve represents the time dependent signal transmitted through the cavity, setting $L$= 7800 μm. Peaks from P1 to P5 include the main transmitted pulse and those triggered by the FP mechanism within the fixed Si plate, therefore they do not depend on L. Peaks from P6 to P8 are also grouped, they arise from internal FP reflections occurring within the cavity and can be controlled using the movable plate.

Both clusters P1-P5 and P6-P8 display on average a temporal delay between adjacent peaks $\tau = 2d\,n_s/c \cong 9\,ps$ (Fig.4(a)) [21], corresponding to the signal round trip within each Si substrate having a measured refractive index $n_s = 3.4$ [21]. The peak P6, originating from the main signal reflections inside the medium (air) bounded by the VO$_2$ layers, can be used to estimate the cavity length using $L = c\,\tau'/2$, where $\tau'$ is the time interval between P1 and P6. According to the main peak width $\delta t = 0.1\,ps$ we can estimate an uncertainty on the cavity length $\delta L = \pm 30 \mu m$. It is worth to mention that there is another cluster of peaks produced by the internal reflections of P2 and further delayed by $2\tau'$, well beyond our range of investigation.

It is interesting to discuss the reason the time profile of peaks P7 and P8 in Fig. 4(a) is inverted in comparison with all other ones. Neglecting losses (the imaginary component) in the Si complex refractive index, one can write the reflection coefficient for a signal propagating in the air towards the plate as $r_{air,Si} = \frac{n_s - n_{air}}{n_{air} + n_s} = -r_{Si,air}$. P6 originates from the beam reflection on interfaces I3 and I2, both providing a positive coefficient $r_{air,Si} \approx 0.55$. P7 is again the result of two reflections, the first one coming from interface I4, with a negative coefficient $r_{Si,air}$, the second one from interface I2 with a positive coefficient $r_{air,Si}$. Following the same argument, P8 does not change in sign with respect to P7 since the beam back-and-forth reflections inside the silicon plate provides an overall reflection coefficient $r_{Si,air}^2 > 0$.

Varying the cavity length, one can easily change $\tau'$ and therefore the relative position of the cluster P6-P8 on the time axis with no other effects on the features discussed above. This is clearly displayed by the red curve in Fig. 4(a), representing the time dependent signal transmitted through the PPFPC setting $L = 6800$ μm. Therefore, reducing $L$ via the movable Si plate the peaks P6, P7 and P8 start overlapping the cluster P1-P5, implying that the time delay introduced by the cavity FP reflections becomes comparable with the one produced by the same mechanism inside the fixed Si plate.

In Figs. 5(a) and 5(c) it is shown how the electric field intensity $E_c(t)$ evolves for different values of the cavity length at T = 50 °C and T = 90 °C, respectively. In Figs. 5(b) and 5(d) the modulus of the corresponding transmission functions $|\tilde{T}(\omega)|$ is displayed. The experimental complex transmission is evaluated by the ratio $\tilde{T}(\omega) = \tilde{E}_T(\omega)/\tilde{E}_r(\omega)$, where $\tilde{E}_r(\omega) = e^{i\omega(L+2d)/c}$ is the reference signal propagating in air along the overall length of the device. Curves are vertically shifted for the sake of clarity. Red curves in panels (b) and (d) represent the $|\tilde{T}|^2$ frequency dependence relatively to a single Si substrate, to show the extrapolated contribution of the transmission through the two plates only.

In Fig. 5(c), although signals are about 1/7 smaller than the T < T$_c$ case, the damping mechanism is reduced because the decay of peaks P1-P5 is not so prominent. This happens because the reflection coefficients of waves bouncing within the Si substrates are larger at higher temperatures. Since the change in amplitude between peaks P1 and P2 is equal to $r_{n_s,n_c}^2$ at 50 °C and $r_{n_s,n_c} r_{n_s,n_v}$ at 90 °C, and assuming a value for the refractive index (real part) of VO$_2$ $n_v = 25$ [17], one yields $\left[\frac{(n_s - n_c)}{(n_c + n_s)}\right]^2_{n_c=1} = 0.29$ in the OFF mode and $\left[\frac{(n_s - n_c)}{(n_c + n_s)} \frac{(n_v - n_s)}{(n_v + n_s)}\right]_{n_c=1} = 0.41$ in the ON mode.

Above T$_c$ the metal-like behavior of VO$_2$ promotes THz waves to oscillate mostly in the cavity and not in the Si plates, since FP reflections within a single Si/VO$_2$ substrate are strongly suppressed. As shown in Fig.(4c), decreasing $L$ $E_c(t)$ undergoes a more pronounced transformation tending to assume the time dependent signal of a resonating process. The mechanism underlying the resonant features in the frequency response of the PPFPC is essentially produced by the superposition of cavity and plate FP modes, as it will be explained more in detail in the next Section. This is clearly shown in Fig. 5(d), where the evolution of transmission $|\tilde{T}|$ is reported as a function of $L$. As long as the cavity length is large enough to keep the FP mode periodicity in the air cavity $\Delta f_{FP_c}$ smaller than the corresponding value in the VO$_2$/Si substrate, $\Delta f_{FP_s}$, the transmission displays a number of resonances basically corresponding to the FP modes of a single plate. The red curve in the figure, representing $|\tilde{T}|^2$ of a single Si/VO$_2$ plate, helps to better discern the frequency position of the related peaks. Actually, cavity modes are mostly hidden by the resonance curve convolution due to the fast modulation of the plate modes. From the lower to the upper black curves, $L$ is reduced until the effective optical length of the cavity fulfils $Ln_c \leq dn_s = 1360$ μm, which implies $\Delta f_{FP_c} > \Delta f_{FP_s}$. Under these conditions, the electrodynamic regime changes and in the PPFPC response only a few peaks survive, corresponding to the frequency positions where there is a complete overlapping of cavity and plate FP modes.

## 5. The transmission model

The signal transmitted across the PPFPC can be modelled using the appropriate combination of Fresnel coefficients and propagation factors, as reported for instance in [20] [22]. An electric field $\tilde{E}_j(\omega)$ impinging at each interface is described considering two media with refractive indexes $\tilde{n}_j$ and $\tilde{n}_{j+1}$ and Fresnel coefficients $\tilde{r}_{j,j+1} = (\tilde{n}_{j+1} - \tilde{n}_j)/(\tilde{n}_{j+1} + \tilde{n}_j)$ and $\tilde{t}_{j,j+1} = 2\tilde{n}_j/(\tilde{n}_{j+1} + \tilde{n}_j)$, in reflection and transmission respectively. The $j$-th component of the transmitted field passes through the ($j+1$)-th medium with refractive index $\tilde{n}_{j+1}$ and thickness $d_{j+1}$, gaining a propagation factor $\tilde{\mathcal{P}}_{j+1} = e^{i\omega d_{j+1}\tilde{n}_{j+1}/c}$. Then, the total electric field $\tilde{E}_T(\omega)$ is obtained summing up the contributions over all possible paths that THz waves can walk to get across the PPFPC.

Therefore, the general expression for the signal transmission through a multiple interface sample consisting of $N$ optical paths can be written in the following way:

$$\tilde{T} = \sum_{j=1}^{N} \left( \prod_{l,m} \tilde{t}_{l,m} e^{i\tilde{n}_m \delta_m \frac{\omega}{c}} \right)_j \widetilde{FP}_j \tag{3}$$

where indices $l$ and $m$ label four media: $a$, air; $c$, cavity; $s$, silicon; and $v$, vanadium dioxide. In eq. (3) only the terms describing the lowest loss paths will be considered, that account for the FP mechanisms either developing within the same medium, characterized by two consecutive interfaces $I_j$ and $I_{j+1}$, or involving reflections between interfaces separating two adjacent media, that is $I_j$ and $I_{j+2}$.

In Figs. 6 and 7 a sketch of the main terms contributing to the transmission expression given in eq. (3), when VO$_2$ is in the insulating or the metal-like phase respectively, is shown. Each blue arrow in the figures represents a low loss optical path term in eq. (3).

Below $T_c$, since VO$_2$ is in the insulating state and therefore mostly transparent to the electric field, signal transmission is calculated considering the FP mechanisms in between four couples of interfaces only: I1-I2, I2-I3, I3-I4 and I1-I3, so that there are four multiple optical paths governing the electrodynamic response (Fig. 6). Above $T_c$, the contribution to the cavity internal reflections given by the VO$_2$ layers acting as reflecting mirrors – and therefore six interfaces – should be considered. In such a case, the preeminent transmission/reflection multiple paths increase to eight (Fig. 7).

Here, as an example, we report the expressions of the $j$-th transmission term describing the electric field reflected inside the air cavity below T$_c$ (Fig. 6(c)):

$$\tilde{T}_j = \tilde{t}_{a,s} e^{i\omega d(\tilde{n}_s - 1)/c} \tilde{t}_{s,c} e^{i\omega L n_c/c} \tilde{r}_{cs} e^{i\omega L n_c/c} \tilde{r}_{c,s} \tilde{t}_{c,s} e^{i\omega d(\tilde{n}_s - 1)/c} \tilde{t}_{s,a} \widetilde{FP}_j \tag{4}$$

where $\widetilde{FP}_j = 1/(1 - \tilde{r}_{c,s}^2 e^{i\omega 2L n_c/c})$,

and above T$_c$ (Fig. 7(c)):

$$\tilde{T}_j = \tilde{t}_{a,s} e^{i\omega d(\tilde{n}_s - 1)/c} \tilde{t}_{s,v} e^{i\omega \delta \tilde{n}_v/c} \tilde{t}_{v,c} \tilde{r}_{c,v} e^{i\omega L n_c/c} \tilde{r}_{c,v} \tilde{t}_{c,v} e^{i\omega \delta \tilde{n}_v/c} \tilde{t}_{v,s} e^{i\omega d(\tilde{n}_s - 1)/c} \tilde{t}_{s,a} \widetilde{FP} \tag{5}$$

where $\widetilde{FP}_j = 1/(1 - \tilde{r}_{c,v}^2 e^{i\omega 2L n_c/c})$.

Once each contribution to the signal is evaluated, then the general expression $\tilde{T} = \sum_{j=1}^{N} \tilde{T}_j$ is compared with the experimental transmission. In Fig. 8(a)[(c)] and 8(b)[(d)] the transmission model in eq. (3) is verified setting $L$=640 μm [$L$=240 μm] for insulating and metal-like VO$_2$ respectively. It is straightforward to observe, as highlighted by the black arrows in the figures, that transmission peaks due to the FP mechanism in the air cavity are clearly visible both below and above T$_c$, despite the background provided by the signal reflections in the Si plates.

The FP resonances inside the whole structure can be characterized using the quality factor $Q_l = \frac{f_l}{\Delta f_l(-3\,dB)}$, where $f_l$ is the resonance frequency and $\Delta f_l$ (-3 dB) is the Full Width Half Maximum (FWHM) frequency interval.

At T = 50 °C we observe an enhancement of the quality factor of both kind of resonances as the frequency increases: $Q_l$ values for the Si/VO$_2$ plate and for the cavity volume vary from 10 to 50 in the range 0.2 – 1.5 THz. At T = 90 °C instead the lossy behavior of the VO$_2$ films tends to suppress the FP mechanism in the Si plates, producing at the same time approximately a two-fold increase in the $Q_l$ values, associated with the resonances inside the air cavity because of the better field confinement.

Resorting to the FP term expressed by eq. (2), this 2-fold increase is due to the drastic change in the VO$_2$ refractive index, so that the inequalities $\frac{\tilde{n}_v}{\tilde{n}_s}, \frac{\tilde{n}_v}{\tilde{n}_c} \gg 1$ hold, which in turn implies $\tilde{r}_{c,v} \to 1$. The resulting response corresponds to a series of delta-like functions driven by the FP function of the cavity: $\lim_{\tilde{n}_v \gg \tilde{n}_s, \tilde{n}_c} \widetilde{FP}_j \to 1/[1 - e^{i\omega 2Ln_c/c}]$.

The experimental evidence shown in Fig. 5(d) proves that when the VO$_2$ films are in the conducting phase, the $|\tilde{T}(\omega)|$ behavior is governed by the distribution of maxima given by the Fabry-Perot term related to the air cavity and described in eq. (5). Since the corresponding term describing the signal reflections in the Si/ VO$_2$ parallel plate shows on the opposite minima very close to zero, then robust transmission maxima can be simply realized by the frequency superposition of the two distinct maxima.

Frequency peaks in the FP terms follow the rule $f_{max} = l\,c/2n_x\delta$, where $n_x\delta$ is the effective optical length of the medium and $l = \pm 1, \pm 2 \ldots$ Hence maxima in the Si/VO$_2$ plate (p) and in the air cavity (c) distribute according to:

$$f_{l_p} = l_p\,c/2n'_s(d+t) \quad (6a)$$

$$f_{l_c} = l_c\,c/2n_c L \quad (6b)$$

where $n'_s = (n_s d + n_v t)/(d + t)$ represents the effective refractive index of the plate [23], that accounts also for the reduction in the periodicity of the FP oscillations in the plate at $T = 90\,°C$ (see Figs. 8(b) and (d)).

Maxima superposition ($f_{max}^p = f_{max}^L$) implies that $l_p n_c L = l_c n'_s(d+t)$. Therefore, as long as $n_c L > n'_s(d+t)$, PPFPC maxima can be obtained with ease since $l_p < l_c$, calling for the existence of some cavity order $l_c$ matching a lower order of the plate $l_p$. In case $n_c L < n'_s(d+t)$, the opposite condition $l_p > l_c$ is more difficult to fulfil experimentally, since the first cavity order ($l_c = 1$) will match a plate FP peak for $l_p > 1$ only, at the expense of an increasingly reduced transmission.

Hence, when VO$_2$ undergoes the phase transition, in the observed transmission $|\tilde{T}|$ most of the peaks disappear and the response becomes "discretized" (fig. 8), with only few resonances left coming from the superposition between FP modes in the cavity and in the plates.

For example, taking into account that $\frac{c}{2dn_s} \sim 0.11\,THz$, the peaks shown in Fig. 5(d) for $L = 730\,\mu m$ placed at about $0.4\,THz$ and $0.6\,THz$ correspond to the orders of cavity and plate modes $(l_c, l_p) = (2,4)$ and $(l_c, l_p) = (3,6)$ respectively. Varying the cavity length to $L = 240\mu m$, two resonances appear at about $0.62\,THz$ and $1.28\,THz$, produced by the mode matching described by the orders (1,6) and (2,13) respectively.

### 6. Cavity sensitivity

The FP resonance superposition mechanism combined with the variable length of the PPFPC can be exploited to realize a high sensitivity sensor that detects tiny changes in the surrounding environment. Inserting a sample (a microfluidic channel, a thin film, a liquid layer, etc.) inside the cavity, $L$ can be varied and optimized to obtain the best comb-like spectrum, featuring resonances with the highest possible peak amplitude along with a reasonable Free Spectral Range $FSR_{l_c} = f_{l_c+1} - f_{l_c} = c/2Ln_c$ [8]. The minimum requirement for the frequency separation is set by the empirical condition $FSR_{l_c} > \Delta f_{l_c}(-3dB)$, the FWHM of the single cavity peak.

Sensor sensitivity $S_{l_c}$ can be evaluated looking at the frequency shift $\delta f_{l_c}$ of a single resonance of order $l_c$ induced by a change $\delta n_c$ in the cavity refractive index. Using eq. (6b) we obtain:

$$S_{l_c} = \frac{\delta f_{l_c}}{\delta n_c} = -\frac{c\, l_c}{2L\, n_c^2}. \qquad (7)$$

Therefore, a high sensitivity can be obtained working with a cavity having a small length in a low refractive index background. From eq. (6b), eq. (7) can be furtherly modified in $\frac{\delta f_{l_c}}{f_{l_c}} = -\frac{\delta n_c}{n_c}$. Using the angular Rayleigh criterium [24], we can estimate the minimum detectable frequency shift as the half resonance width $\delta f_{l_c} = \Delta f_{l_c}/2$. Thus, using the experimentally measured value of the resonance quality factor $Q_{l_c} = \frac{f_{l_c}}{\Delta f_{l_c}}$, we can evaluate the smallest detectable relative variation of the cavity refractive index as

$$\frac{\delta n_c}{n_c} = \frac{1}{2Q_{l_c}}. \qquad (8)$$

For $Q_{l_c} = 95$, from eq. (8) one yields $\delta n_c/n_c = 0.005$, implying that a variation rather less than 1% can be measured looking at the resonance shift of the vanadium-based PPFPC.

To provide a more quantitative information on the cavity sensitivity when the temperature of the plates is T = 90 °C, we have also performed full-wave simulations of the cavity response using a commercial EM code (CST Microwave Studio®), evaluating the dependence of the resonance frequency position as a function of $n_c$. In Fig. 9 the shift of the peaks for $l_c = 1, 2$ extracted from full wave simulations of the cavity for $L = 240\mu m$ (open symbols), are compared with the results obtained through the analytical model (full symbols) varying the refractive index in the range 1÷1.4. A linear fit of the two peak frequency dependence on $n_c$ returns as sensitivity the $S_1 = -0.43\ THz/RIU$ and $S_2 = -0.87\ THz/RIU$ values, which represent, to the best of our knowledge, the highest values available in literature for THz sensors [25] [26] [27]. As expected, sensitivity improves increasing the resonance mode order.

Summarizing, we have reported a study on the sensing properties of a Parallel Plate Fabry Perot Cavity having a tunable length and VO₂ layers as mirrors, that can be thermally switched from an insulating to a metal-like behavior passing through a phase transition temperature $T_c$. The control of the VO₂ state enables to overcome the problem of the classical open cavities realized with metallic mirrors, that present very low transmission spectra. The proposed device in fact is designed so that its comb-like spectrum can be primarily adjusted and optimized, first by keeping the VO₂ in the insulating state and then inducing the transition in the metal-like state during the sensing operations.

We measured the transmission of THz waves for different cavity lengths in the two states of the vanadium dioxide and compared the results with an analytical model, that accounts for the relevant contributions to the electric field provided by the FP resonances within the different media marked by the metallic and dielectric interfaces of the device. We found that the quality factor of the overall cavity shows a 2-fold increase when the VO₂ films undergoes the phase transition, therefore the PPFPC can be optimally exploited to detect small variations in the refractive index of the cavity itself. The conceived system shows a very high sensitivity to tiny changes in the surrounding environment, suggesting its use as a reliable sensor for applications like monitoring of water contamination or air pollution, and for quality control and in-line verification in the agri-food industry [28].

**Acknowledgments**

A.A. and G.P.P. gratefully acknowledges the support of Istituto Nazionale di Fisica Nucleare (INFN) under the project "TERA". A.L.P. and G.M. thank the Bio-nanotech Research and Innovation Tower (BRIT) laboratory of the University of Catania (Grant no. PONa3_00136 financed by the MIUR) for the Smartlab diffractometer facility.

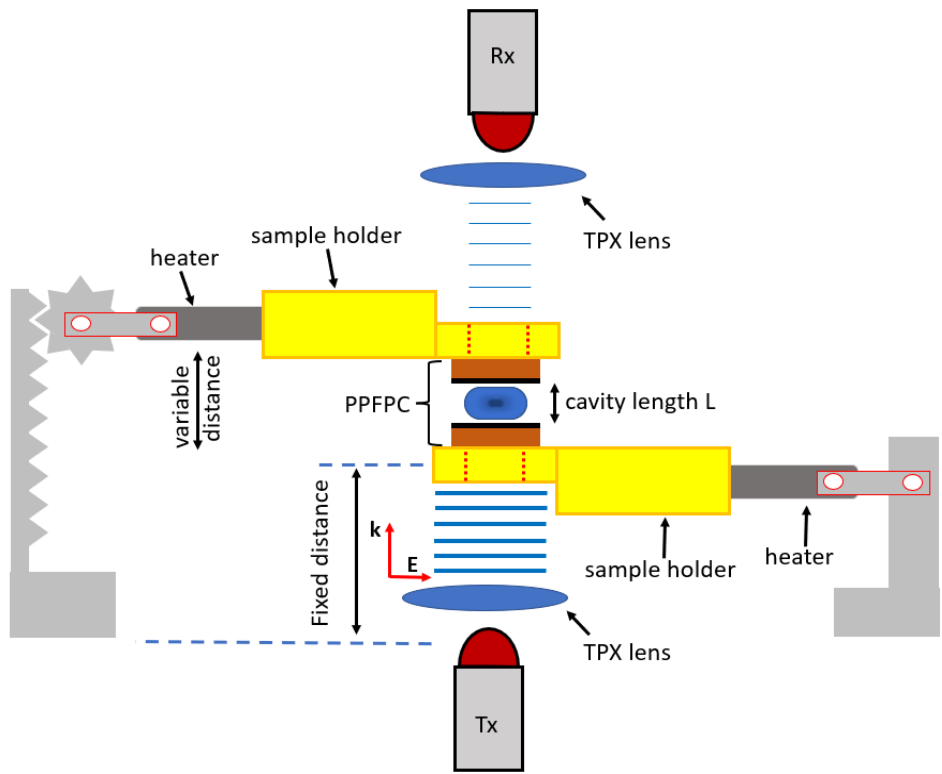

*Figure 1: Sketch of the optical setup. Tx and Rx are the emitter and detector respectively. Two TPX lenses are used to collimate the beam and ensure a transmitted THz signal as close as possible to a planar wave. One sample holder with heater has a fixed position, the other one can be moved using a motorized linear stage.*

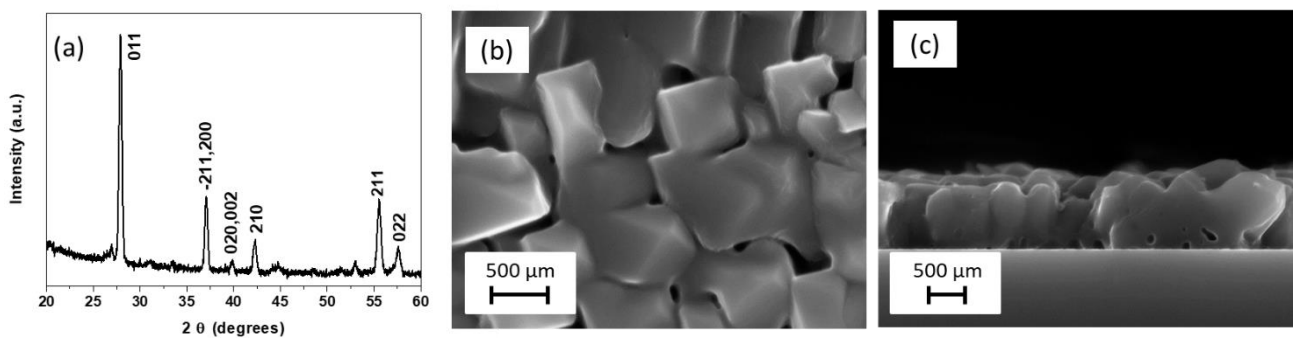

*Figure 2. (a) X-ray diffraction pattern, and FE-SEM plan-view (b) and cross-sectional (c) images of the $VO_2$ film.*

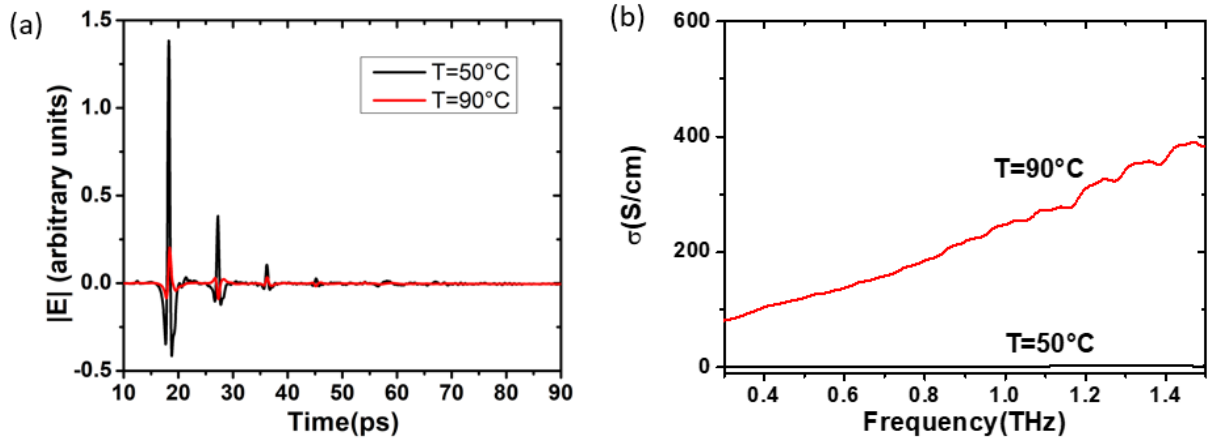

Figure 3: (a) time dependent signal and (b) conductivity of the VO$_2$ film on Si at temperatures lower and higher than T$_c$.

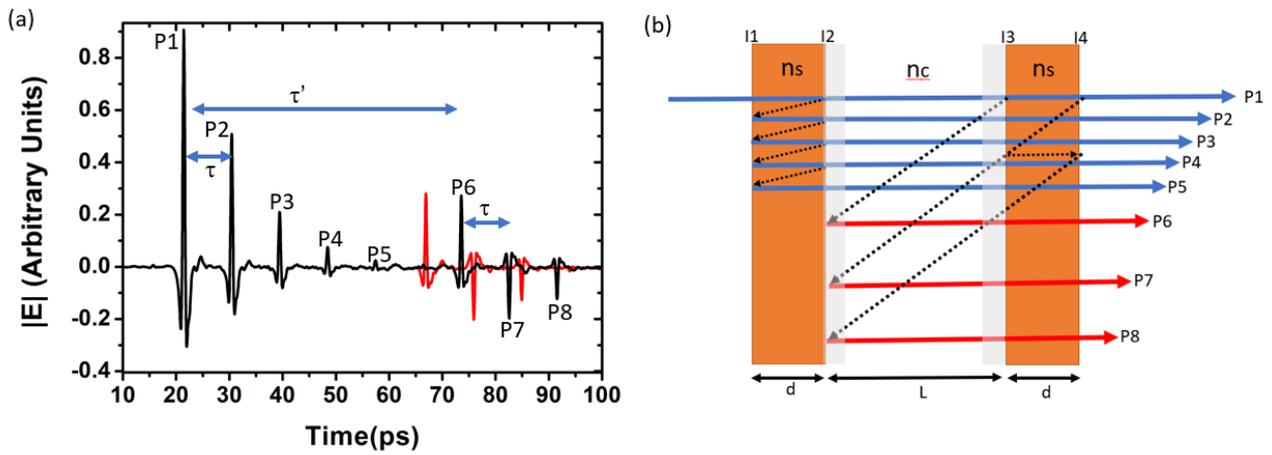

Figure 4: (a) The black curve shows the time dependent electric field signal transmitted across the PPFP setting L = 7800 µm. The red dashed curve represents the same signal for L = 6800 µm. (b) Sketch of the system at T = 50 °C with indication of the main transmitted pulse P1 and its reflections as labelled in panel (a). Blue (red) lines refer to the multiple reflections of the main signal inside the Si plate (within the air cavity). Labels I1-I4 represent the system interfaces.

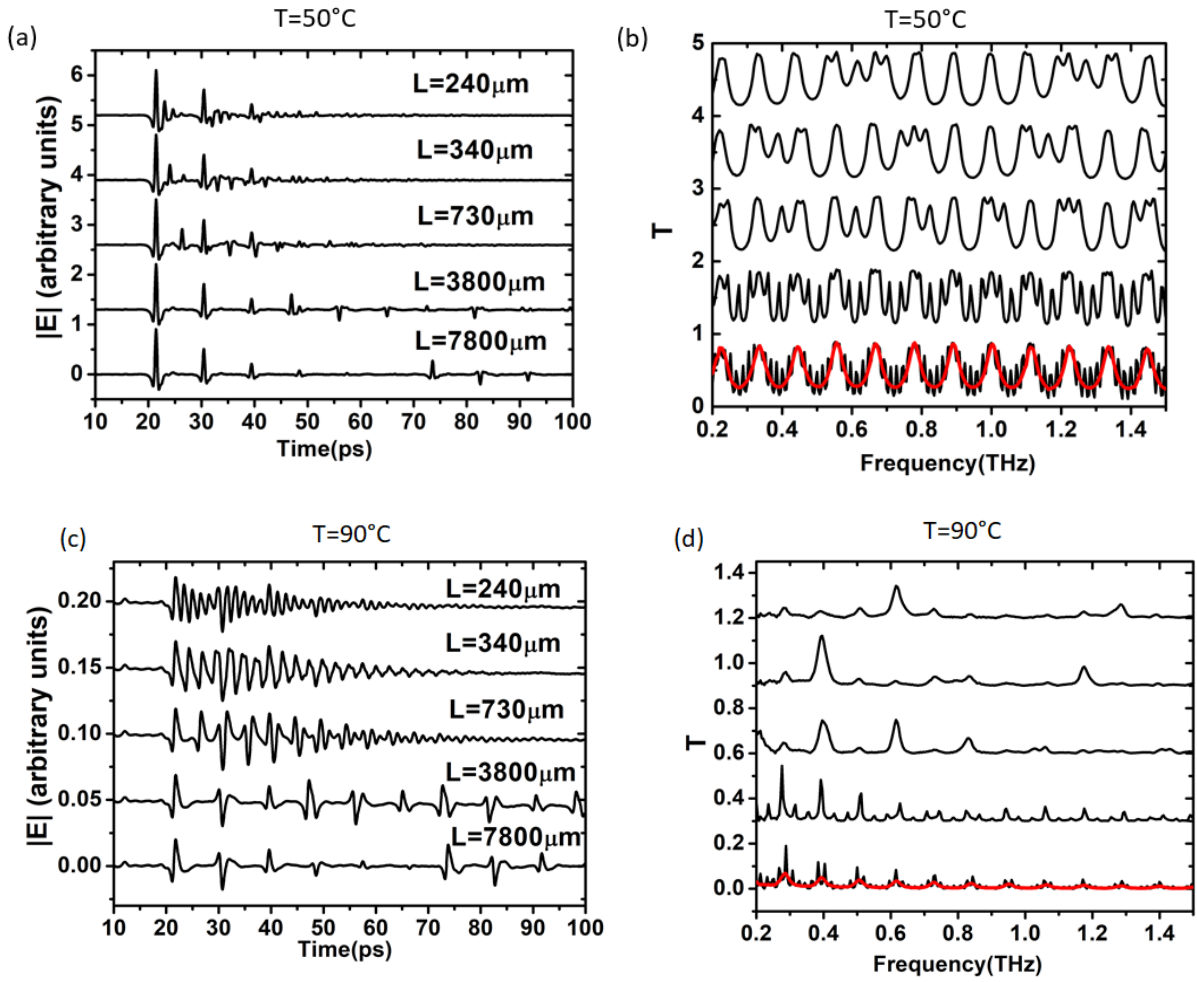

Figure 5: (a), (c) time dependent electric field signal and (b), (d) transmission through the PPFPC as a function of frequency for different values of L at T = 50 °C and T = 90 °C respectively. Black curves are vertically shifted to enhance the comparison. Red curves in panels (b) and (d) represent the $T^2$ frequency dependence relatively to a single plate of the cavity.

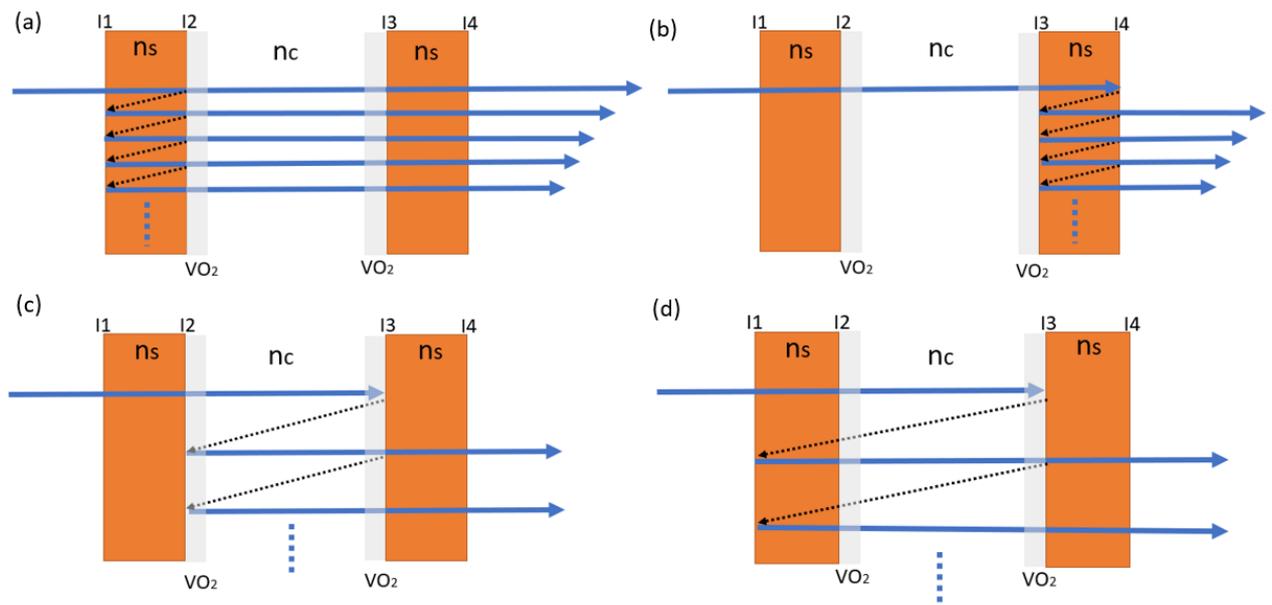

*Figure 6: Major transmission/reflection paths contributing to the transmission function T when the VO$_2$ layers are in the insulating phase (T = 50 °C).*

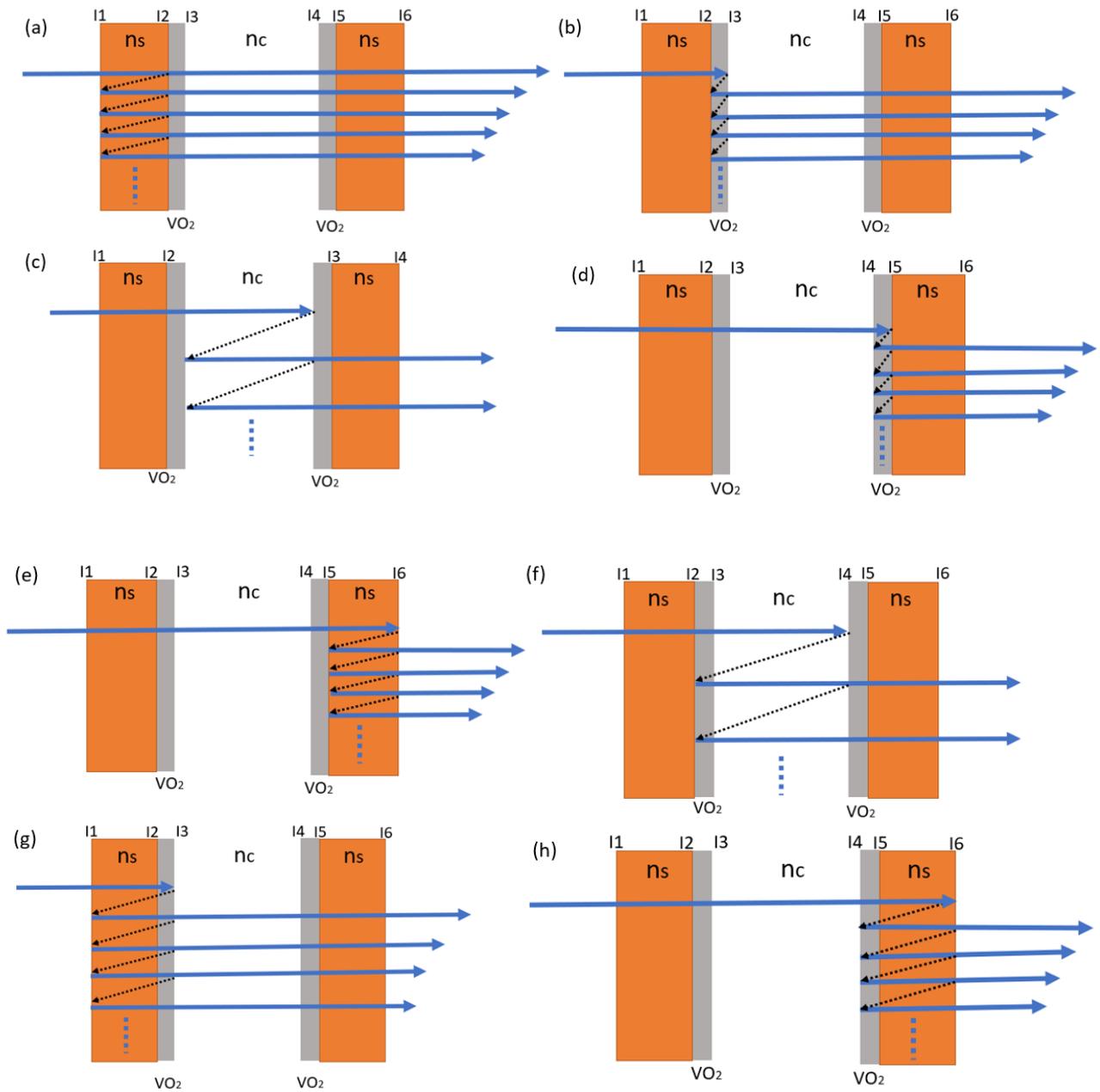

*Figure 7: Major transmission/reflection paths contributing to the transmission function T when the VO₂ layers are in the conducting phase (T = 90 °C).*

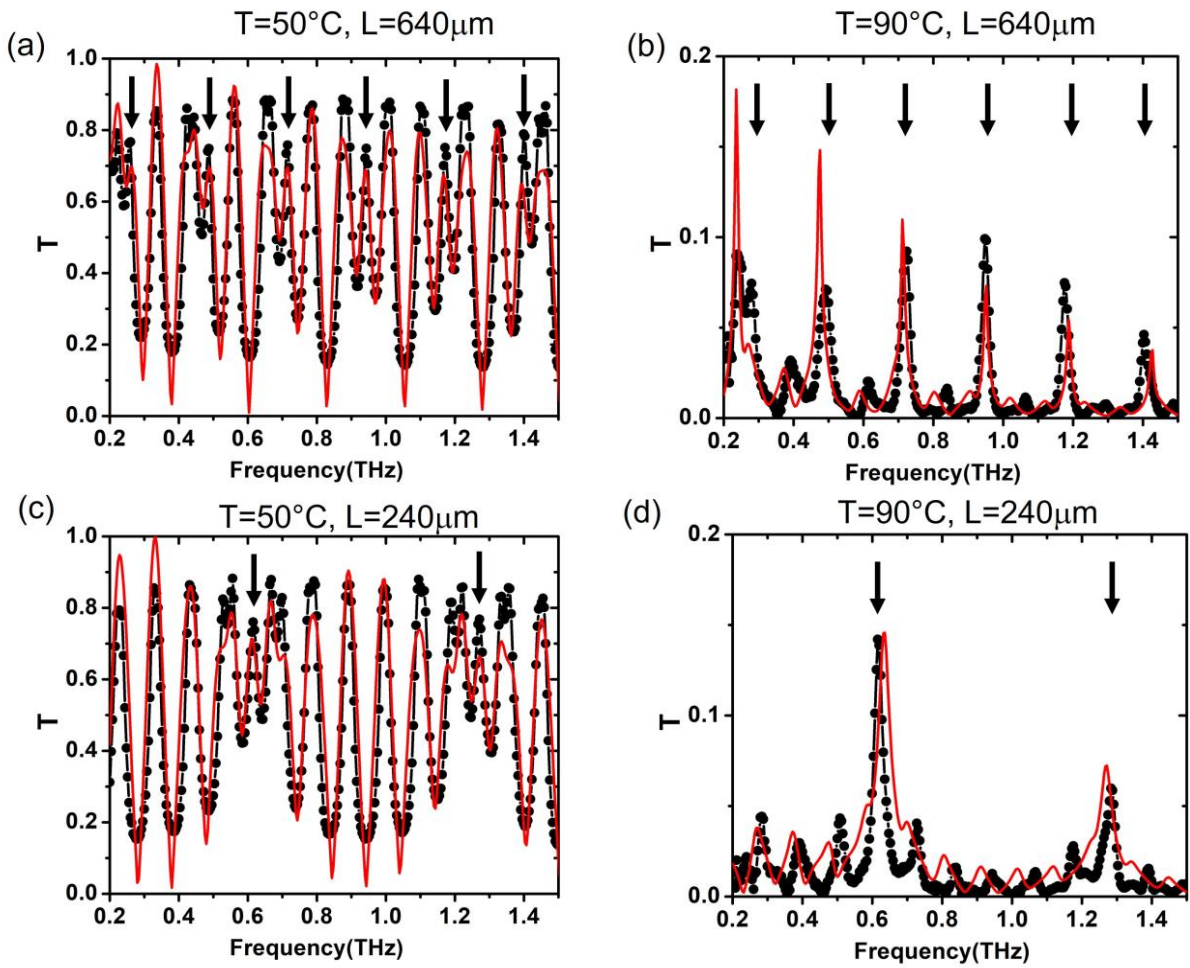

*Figure 8: Comparison between the experimental transmission T (black dots) through the PPFPC and the model (red curve) expressed by eq. (3) for L = 640 µm and L = 240 µm, below and above $T_c$. Black arrows indicate the resonances due to the FP mechanism in the cavity.*

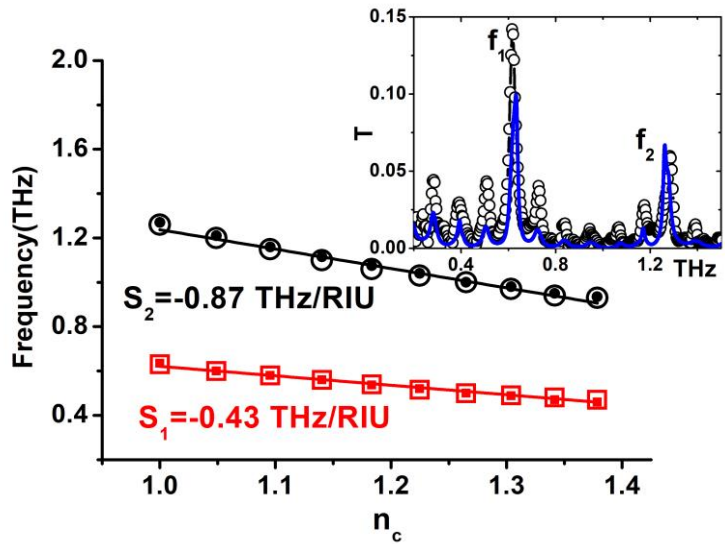

*Figure 9: Dependence of the resonance frequency on the refractive index for the first two order FP modes in the cavity. Open symbols represent full-wave simulations whereas full symbols describe the results by using the analytical model. Continuous curves represent the best linear fits of the resonance shifts as a function of the refractive index of the cavity. Inset: black dots represent the measured transmission for L = 240 mm and T = 90 °C whereas the blue curve is the full-wave simulation.*